# Environmental Instability and Degradation of Single- and Few-Layer $WTe_2$ Nanosheets in Ambient Conditions

Fan Ye[1†], Jaesung Lee[1†], Jin Hu[2], Zhiqiang Mao[2], Jiang Wei[2], Philip X.-L. Feng[1*]

[1]*Department of Electrical Engineering & Computer Science, Case School of Engineering, Case Western Reserve University, 10900 Euclid Avenue, Cleveland, OH 44106, USA*

[2]*Department of Physics and Engineering Physics, Tulane University, New Orleans, LA 70118, USA*

## Abstract

**Since the discovery of large, non-saturating magnetoresistance in bulk $WTe_2$ which allows microexfoliation, single- and few-layer $WTe_2$ crystals have attracted increasing interests. However, as it mentioned in existing studies, $WTe_2$ flakes appear to degrade in ambient conditions. Here we report experimental observations of saturating degradation in few-layer $WTe_2$ through Raman spectroscopy characterization and careful monitoring of the degradation of single-, bi- and tri-layer (1L, 2L & 3L) $WTe_2$ over long time. Raman peak intensity decreases during $WTe_2$ degradation and 1L flakes degrade faster than 2L and 3L flakes. The relatively faster degradation in 1L $WTe_2$ could be attributed to low energy barrier of oxygen reaction with $WTe_2$. We further investigate the degradation mechanisms of $WTe_2$ using XPS and AES and find that oxidation of Te and W atoms is the main reason of $WTe_2$ degradation. In addition, we observe oxidation occurs only in the depth of 0.5nm near the surface, and the oxidized $WTe_2$ surface could help prevent inner layers from further degradation.**

[†]Equally contributed authors. [*]Corresponding Author. Email: philip.feng@case.edu.

# 1. Introduction

Tungsten ditelluride ($WTe_2$) has recently been attracting significant and increasing interests because of the discovery of large, non-saturating magnetoresistance[1] in bulk $WTe_2$ and the prediction that strained single-layer $WTe_2$ can exhibit two-dimensional (2D) topological transitions[2], which may not be easily assessable in other 2D materials. Under ambient conditions, $WTe_2$ usually stays in its Td phase: $WTe_2$ layers stack in a direct fashion, resulting in a higher-symmetry orthorhombic structure[3]. In the present efforts and processes of fabricating few- and single-layer $WTe_2$ devices, one noticeable obstacle is its surface degradation in ambient conditions[4,5,6]. For example, in the study of metal-to-insulator transition in few-layer $WTe_2$ field effect transistors (FETs) at low temperature, it is suggested that the observed transition could be attributed to the increasing disorder of $WTe_2$ lattices caused by degradation[4]. In addition, while it is already predicted that spin Hall effect could be observed in strained single-layer $WTe_2$[2], demonstration of functioning single-layer $WTe_2$ devices is yet to be achieved, despite recent efforts and attempts[3]. The susceptibility to environment and degradation in ambient could have been a major limitation. Also importantly, whether the large non-saturating magnetoresistance could be observed in few- and single-layer $WTe_2$ is still unknown. All these aforementioned intriguing possibilities demand high-quality and reliable few- and single-layer devices. Though degradation of few-layer $WTe_2$ has been observed and hypothesized in previous experiments [3,4,5], the observed data on environment effects have been scattered, a quantitative, deep understanding of degradation behavior is still lacking, and would be desirable and helpful for the establishment of functional $WTe_2$ devices toward high performance.

In this work, we report a detailed investigation on the degradation of $WTe_2$ using Raman spectroscopy and surface analysis methods, including X-ray photoelectron spectroscopy (XPS) and Auger electro spectroscopy (AES). We find that thinner $WTe_2$ flakes degrade faster than thicker samples in ambient conditions. Moreover, XPS and AES measurements indicate that oxidation is the main origin of $WTe_2$ degradation. Fortunately, degradation occurs mostly on top surface and thus the oxide layer can be quite effective in protecting inner layers from further degradation.

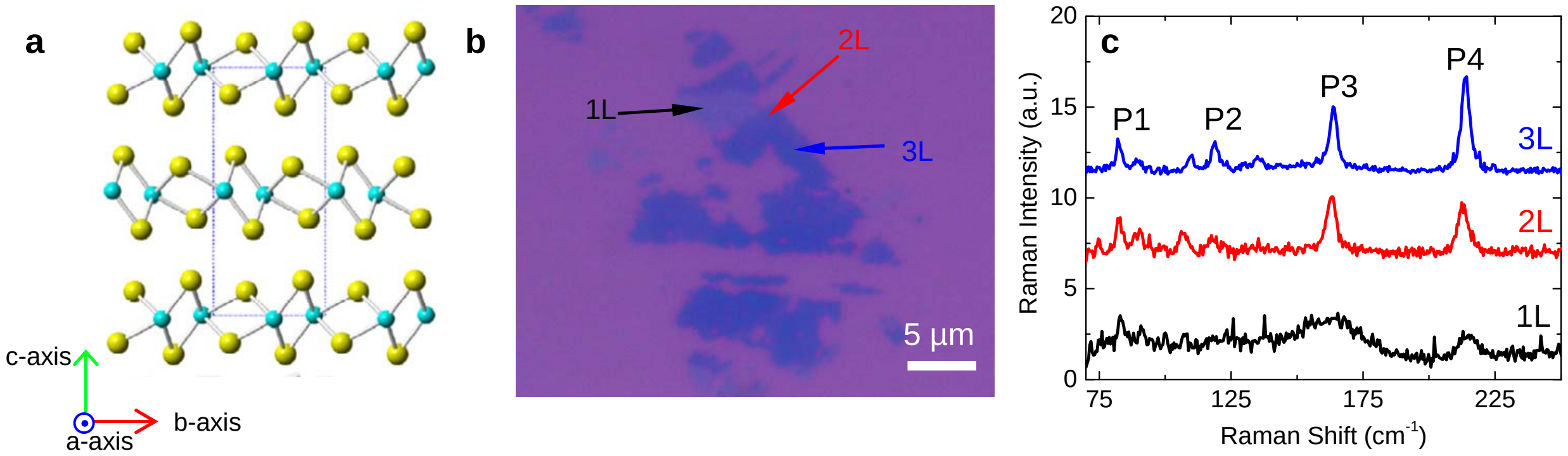

**Figure 1**: $WTe_2$ crystal structure and Raman spectra of single- and few-layer $WTe_2$. (a) Side view of $WTe_2$ structure, blue and yellow spheres represent W and Te atoms respectively. Blue rectangular boxes represents unit cell of $WTe_2$. (b) Optical microscope image of single-layer (1L), bi-layer (2L) and tri-layer (3L) $WTe_2$ flakes. Scale bar: 5μm. (c) Measured Raman spectra of single-layer and few-layer $WTe_2$, where measurement positions correspond to arrows with labels in (b). P1, P2, P3 and P4 are employed for further analyses.

## 2. $WTe_2$ Degradation Studied by Raman Analysis

Figure 1(a) shows the $WTe_2$ crystal structure. The a, b planes form a single-layer $WTe_2$, in which tungsten (W) atoms are sandwiched by two tellurium (Te) atomic sheets. The three nearest Te atoms from each sheet form a triangular pyramid with the W atom, with the two resulting opposing pyramids rotated 180º (along the c-axis) from each other. Compared with some 2D materials that exist in 1T phase such as 1T $MoS_2$, W atoms in $WTe_2$ deviate to their ideal sites, forming distorted octahedral structure. Layer by layer, $WTe_2$ stacks along the c-axis, forming the bulk $WTe_2$ crystal.

Single-, bi- and tri-layer (1L, 2L & 3L) $WTe_2$ flakes are deposited on 290nm $SiO_2$ on the top of Si. Once the samples are prepared, we promptly transfer them into a vacuum chamber and first measure Raman signals in vacuum at room temperature. Subsequently we locate samples out of the vacuum chamber and measure them in ambient air conditions, starting from 5 minutes up to 15 days. During the measurement intervals, samples are stored in ambient conditions.

Figure 1b shows the optical microscopy images of 1L, 2L and 3L $WTe_2$ and Figure 1c shows the Raman spectra of corresponding regions in (b) in vacuum conditions. In our previous study, we have observed total 12 peaks in few-layer $WTe_2$[7]. As number of layers decreases to 3L, many peaks' intensities decrease significantly and only four peaks remain robust due to space group evolution from bulk ($C_{2v}$) to 1L ($C_{2h}$) WTe2[6]. We refer these four peaks as P1, P2, P3 and P4 hereafter (see Figure 1c). Since Raman spectrum is sensitive to crystal quality and structure, we employ these four peaks as indicators in our degradation study.

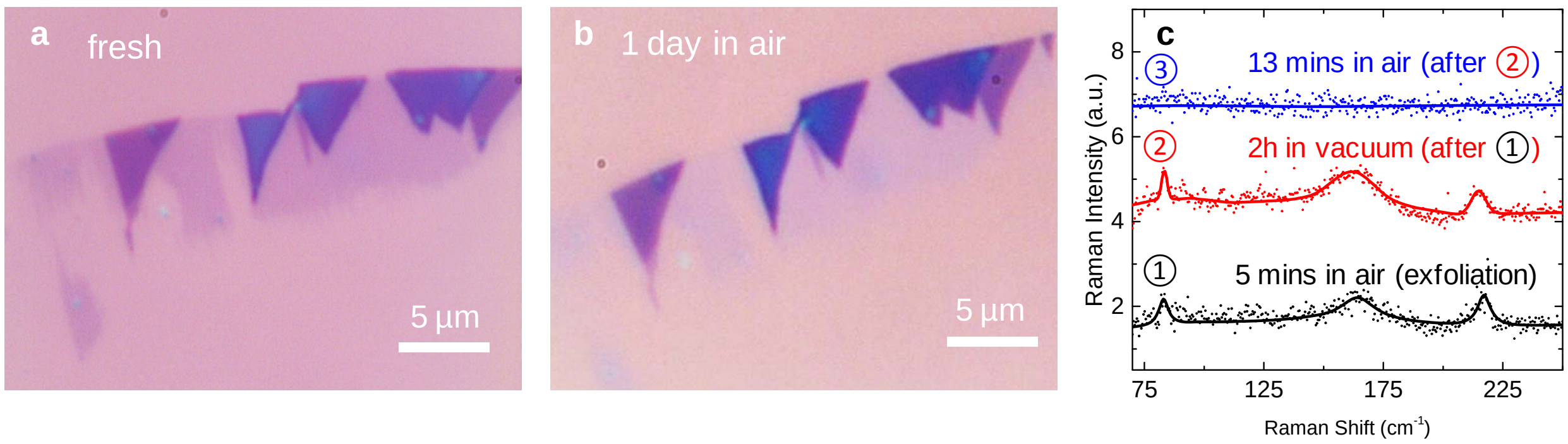

**Figure 2**: Optical images and Raman spectroscopy results of single-layer (1L) $WTe_2$ degradation. Optical microscope image of (a) fresh 1L $WTe_2$ (~5 minutes after exfoliation). (b) $WTe_2$ flake in (a) that exposed to air for 1 day. (c) Raman spectrum of 1L $WTe_2$ over time in vacuum and ambient conditions.

We first investigate degradation of single-layer (1L) $WTe_2$. We store samples in the vacuum chamber immediately after exfoliation and monitor Raman signals over 2 hours (Figure 2c). We find that single-layer $WTe_2$ degrades fairly fast in ambient conditions: from Raman spectrum, it shows some signs of degradation during our exfoliation process that only takes 4–5 minutes. In comparison, single-layer $WTe_2$ is very stable in vacuum conditions: P1, P3 and P4 do not exhibit noticeable peak shift or intensity variation over 2 hours in vacuum (P2 is absent in 1L $WTe_2$ because of transition of space group[6]). After that, flake is exposed to ambient conditions and Raman spectra are measured again. We find that P1, P3, and P4 Raman peaks vanish completely in only 13 minutes (totally 17 to 18 minutes including the exfoliation time of 4–5min) in ambient air. These results indicate that the Td structure of $WTe_2$ is significantly modified during exposure to ambient conditions. In addition, we also notice that the optical contrast of single-layer $WTe_2$ decreases (see Fig. 2) (also mentioned in Ref. [3,4,5]), suggesting that the refractive index of $WTe_2$ is changed. This fairly fast degradation makes it difficult to fabricate high quality single-layer device in ambient conditions. One strategy is to fabricate single-layer $WTe_2$ devices (especially those need electrodes and require longer time) in inert gas and/or vacuum glove boxes, which could prevent $WTe_2$ from degradation.

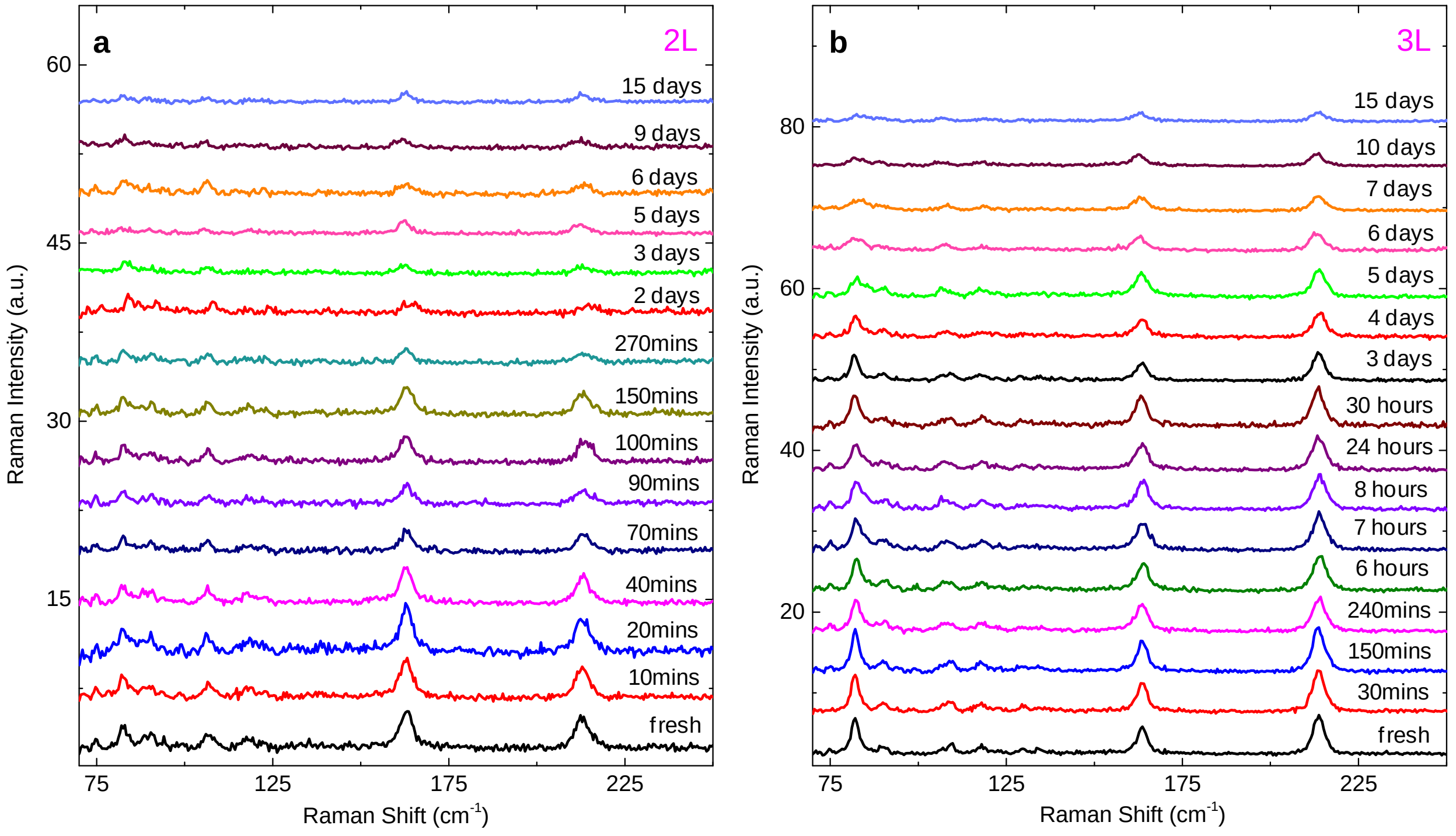

**Figure 3**: Evolution of Raman spectra of bi-layer (2L) and tri-layer (3L) $WTe_2$ in ambient conditions. Measured Raman results of (a) 2L $WTe_2$ and (b) 3L $WTe_2$ during degradation over 15 days.

After investigation on 1L $WTe_2$, we turn our focus to degradation in 2L and 3L $WTe_2$. Figure 3a & b show Raman results of bi-layer and tri-layer $WTe_2$ over 15 days in ambient conditions, respectively. After exposure to ambient conditions the intensities of P1 to P4 in both 2L and 3L decrease significantly, suggesting $WTe_2$ has undergone degradation during this time interval. In 2L $WTe_2$, the intensities of P1 to P4 initially decay and eventually stay stable over exposure in ambient conditions. Similar results are also observed in 3L $WTe_2$. It is worth noting that Raman peaks in 2L and 3L $WTe_2$ still remain clearly detectable after 15 days exposure to ambient conditions, showing much higher environmental stability compared with that of 1L $WTe_2$.

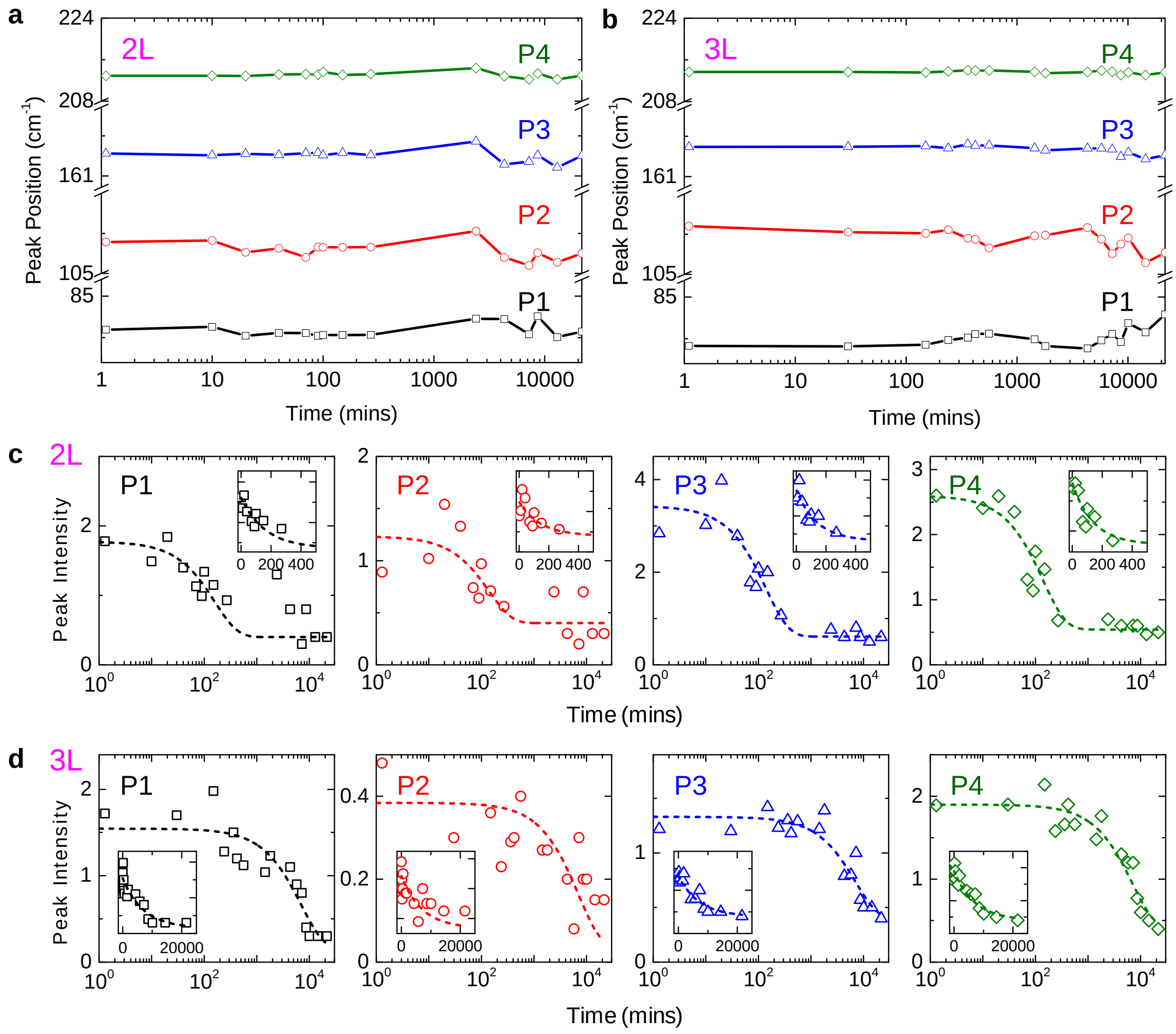

**Figure 4**: Raman position and intensity changes of bilayer and trilayer during $WTe_2$ degradation. P1, P2, P3 & P4 positions of (a) 2L and (b) 3L during degradation. Peak intensities of (c) 2L and (d) 3L over time with exponential decay fitting. Insets show early stage of (c) & (d) with linear *x*-axis.

To further illustrate the degradation behavior in Figure 3, Raman peak positions and peak intensities of 2L and 3L $WTe_2$ over degradation time are shown in Figure 4. In Figure 4 a & b, there are no obvious Raman peak position shifts during the degradation in both 2L and 3L $WTe_2$, which indicates that neither obvious tension variations nor van der Waals (vdW) interactions occur during $WTe_2$ degradation. From Figure 4c & d, we find that intensities of P1 to P4 all decay significantly during degradation. We fit peaks intensity with an exponential decay function of

$$I(t) = I_1 e^{-(t-t_0)/\tau} + I_2 , \quad (1)$$

where $I_1$, is the initial Raman peak intensity, $I_2$ is the intensity after degradation saturation, $t_0$ is the time interval between the very first Raman measurement and the flake exfoliation, $t$ is the time interval between the present Raman measurement and the flake exfoliation, and $\tau$ is the intensity decaying characteristic time, respectively. We find the decay time of 2L $WTe_2$ is $\tau_{2L}$~140–160 minutes, which suggests that the degradation process in 2L $WTe_2$ occur mostly in initial 3 hours of exposure to ambient conditions. From ~3 hours to 15 days, degradation slows down and peak intensities become stable, indicating the saturation of degradation. Similar behavior is also shown in 3L $WTe_2$: exponential decay fitting (Eq. 1) revels decay time of 3L $WTe_2$ $\tau_{3L}$ is $\tau_{3L}$~7000–9000 minutes. Degradation of 3L $WTe_2$ starts gradually after exfoliation and eventually saturates after 2 weeks. These behaviors are different with that of 1L sample. Once 1L flake is deposited on substrate, it undergoes much faster degradation and all Raman signals disappear within 13 minutes. Based on the measured Raman spectra from 2L and 3L flakes, we speculate that degradation in $WTe_2$ mostly occurs at its surface, which is a self-limiting and saturating behavior. Similar degradation tendency has also been observed in $WSe_2$[8,9].

## 3. Element Analysis of Degraded $WTe_2$

To gain further insight into this saturating degradation behavior in $WTe_2$, we investigate surface composition of degraded $WTe_2$ by employing X-ray photoelectron spectroscopy (XPS) and Auger electron spectroscopy (AES). The $WTe_2$ flakes for XPS and AES are exfoliated directly on $SiO_2$ substrates. Compared with Raman spectroscopy, XPS measurements require larger sample due to relatively much big spot size of X-ray beam (~20μm). We choose flake with a thickness of $t$~100nm and a lateral length of $l$~70μm. Since $WTe_2$ is layered material, it is reasonable to believe that the degradation takes place sequentially from outer to inner layers, and thin $WTe_2$ and thick $WTe_2$ flake degrade with same mechanisms. After $WTe_2$ flakes are exfoliated on the $SiO_2$ substrate, they are exposed to ambient conditions for 15 days before the XPS measurements.

Figure 5a & b show the measured XPS spectra of the degraded $WTe_2$ samples. We find clear oxidation signatures of Te and W atoms on the surface of the degraded $WTe_2$[10,11]: appearance of Te-O peaks at binding energy of 576.1 eV and 587.1 eV near Te-3d band (Figure 5a) and W-O peaks at binding energy of 247.4 eV and 260.6 eV near W-4d band (Figure 5b). These XPS observations indicate that the main mechanism of degradation of $WTe_2$ in ambient conditions is oxidation of both W and Te atoms, and $TeO_2$ and $WO_x$ ($2 < x < 3$) are the main oxidation products of $WTe_2$ degradation, which can be estimated from chemical shifts of the Te-3d and W-4d peaks.

To quantitatively investigate the thickness of the oxidized layer on $WTe_2$ surface, we etch away very thin layer (~0.5nm) of the degraded $WTe_2$ by using 3keV $Ar^+$ ion etching in XPS and survey depth-profile of samples. The etching rate is first determined using standard $Ta_2O_5$ sample and then converted to etching rates of $WTe_2$ oxidation products ($TeO_2$ and $WO_x$) using the etching yield values. Interestingly, we find that both W-O and Te-O peaks disappear after etching and the intensity of Te-W bonds increases significantly near Te-3d band and W-4d band (Figure 5a & b). This observation clearly demonstrates that oxidation in ambient conditions only occurs on the top surface of $WTe_2$. Once oxygen atoms are adsorbed on the $WTe_2$ surface, they react with $WTe_2$, and generate oxidation products such as $WO_x$ and $TeO_2$, which passivate the $WTe_2$ surface, preventing oxygen from further diffusing into inside of lattice and protecting inner layer of $WTe_2$. In addition, our observation suggests that $Ar^+$ ion plasma cleaning could be an effective way to refresh surface of oxidized $WTe_2$. We etch one more cycle with same period and acquire XPS signal for the $WTe_2$. After second round of etching, only Te-W bonds exist, which is consistent with the results after first etching. This further indicates that the $TeO_2/WO_x$ layer only exists on top of the degraded surface.

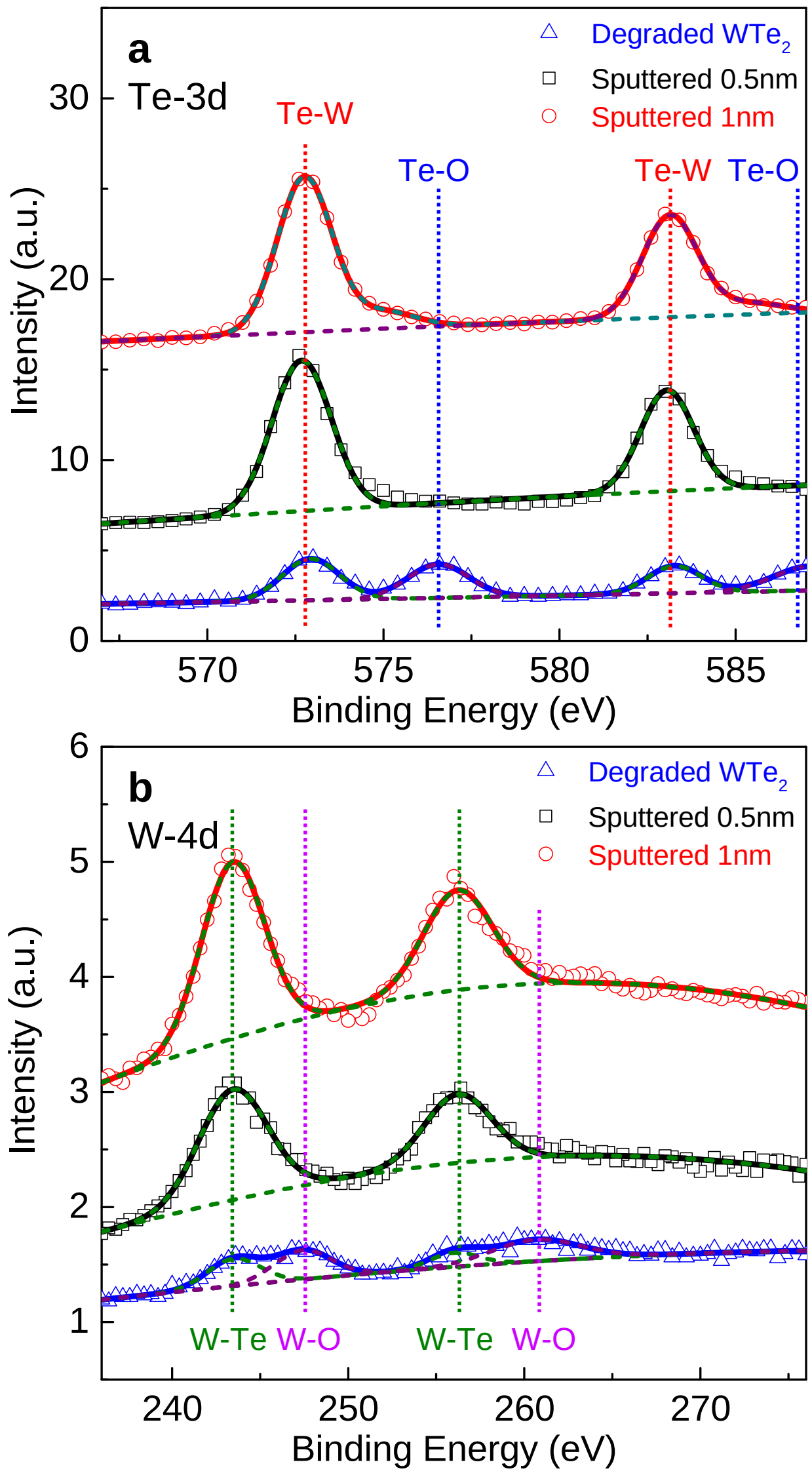

**Figure 5**: XPS spectra of degraded $WTe_2$ before etching (blue line), after 0.5nm etching (black line), and 1nm etching (red line). Measured results near (a) Te-3d (b) W-4d binding energy. Solid lines represent experiments data. Green dash lines are Gaussian fits for W-Te bonds. Purple dash lines are Gaussian fits for Te-O bonds ($TeO_2$) in (a) and W-O ($WO_x$) in (b) respectively. The intensity in (a) and (b) have been offset for clear illustration.

In addition to vertical elemental analysis using XPS, we also perform AES analysis on a $WTe_2$ flake to investigate horizontal oxidation distribution of degraded $WTe_2$. Compared with XPS, AES has higher spatial resolution, which enables in-depth elements mapping on sample surface. Figure 6a & b show the optical and SEM images of sample we used in the AES measurements.

The flake, with a size of *l* ~ 70μm and thickness of *t* ~ 100nm, is stored in ambient conditions for ~15 days before the AES analyses. We first perform elemental mapping on the degraded $WTe_2$ surface. For oxygen mapping, we use peak intensity for Te-O and W-O bands which can be distinguished with oxygen peak from $SiO_2$ due to different chemical shifts. Figure 6c, e and g show oxygen (O), tungsten (W) and tellurium (T) element maps before etching, while Figure 6d, f, and h show their corresponding element maps after etching, respectively. Wide range spectra measured before and after etching are shown in Figure 6i and j. We observe significant oxidation of $WTe_2$ after exposure to ambient conditions, as evident in the oxygen mapping results (Figure 6c), which perfectly agrees with the aforementioned XPS results. After etching ~0.5nm, oxygen area reduces significantly and only small points (small green spots in Figure 6d) exist, revealing nonuniform oxidation in $WTe_2$. Such localized oxidization on $WTe_2$ indicates slightly deeper oxidation of $WTe_2$, which may be generated by defect induced, localized more intense and faster oxidation[12]. There is a folded area at the bottom of the $WTe_2$ flake and it may not be effectively etched and analyzed due to different angles with respect to the $Ar^+$ etching, electron beam gun and detector.

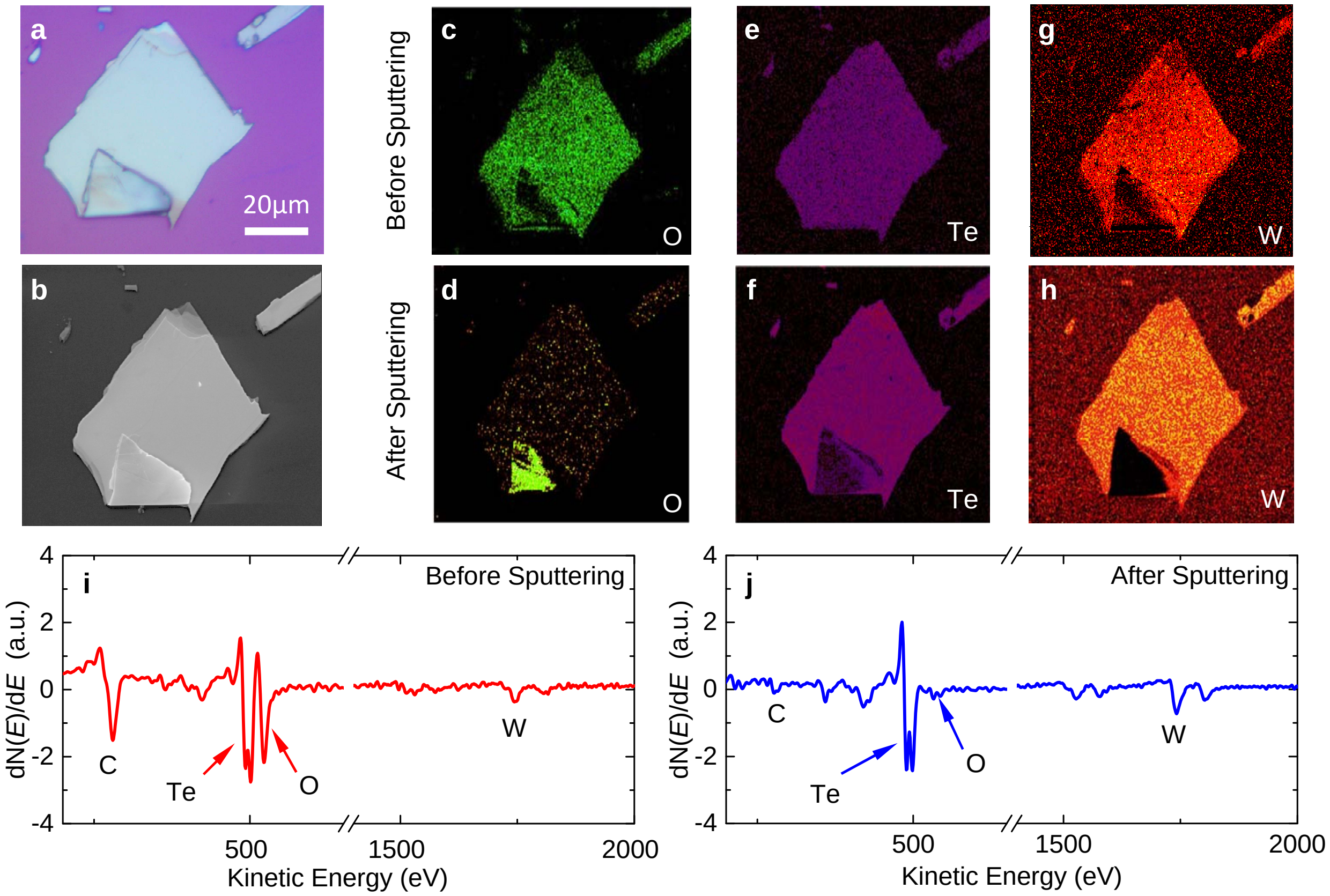

**Figure 6**: AES spectra of degraded $WTe_2$ before and after etching 0.5nm. (a) Optical Image and (b) SEM image of sample used. (c) Oxygen with Te-O and W-O bonds (e) tellurium (g) tungsten (g) elements mapping before etching, and (d) oxygen (f) tellurium (h) tungsten elements mapping results after etching ~0.5nm. Wide range elemental analysis (i) before and (j) after etching.

Based on the above Raman, XPS and AES results, we now focus our discussion on the $WTe_2$ degradation mechanisms. Main driving force of $WT_2$ degradation in ambient conditions is surface oxidation. Oxidation induced degradation has been observed in others TMDCs such as $MoTe_2$[13], $WSe_2$[9,10], and also in black phosphorus[13,14,15]. $WTe_2$ has unique degradation behavior compared with that in other 2D materials: both W and Te atoms are oxidized during exposure to air while only one element is oxidized for $WSe_2$ and $MoTe_2$ (*e.g.*, main oxidation products for $MoTe_2$ and $WSe_2$ are $TeO_2$ and $WO_3$, respectively); 1L $WTe_2$ degrades within 13 minutes in ambient conditions while other 2D materials have exhibited much slower degradation (*e.g.*, 1L $MoTe_2$, $WSe_2$ sustain their properties up to several days in ambient conditions)[13,17]. Such unique degradation in $WTe_2$ is attributed to its low activation energy for oxidation. Once oxygen is adsorbed onto $WTe_2$ surface, it dissociates into two oxygen atoms and react with $WTe_2$ surface, finally leading to surface oxidation of $WTe_2$[16]. Based on theoretical calculations, there is no activation energy for $WTe_2$ degradation (0 eV), which is much lower than those for other 2D materials (*e.g.*, 0.25eV for $MoTe_2$, 0.58eV for $WSe_2$, 0.69eV for black phosphorus), resulting in faster degradation of $WTe_2$[15,17].

According to the discussions above on degradation mechanism of $WTe_2$, it is worth further considering how to avoid degradation for fabricating few- and single-layer $WTe_2$ devices with excellent performance. One common strategy is to use h-BN (hexagonal boron nitride) encapsulation, which can not only avoid or mitigate degradation [14] but also enhance device surface/interface quality and carrier mobility [18,19]. Another method is to mix $WTe_2$ nanosheets with protective polymers, *e.g.*, poly(vinyl alcohol); and this polymer film could prevent nanosheets from degradation, which has been proved to be an effective approach to protecting $WS_2$, $MoTe_2$ and $WTe_2$ crystals[20,21]. Furthermore, in this work, we have found that $Ar^+$ ion plasma etching could refresh degraded $WTe_2$ surface. The above methods, collectively, can be highly constructive to provide guidelines and solutions for fabricating few-, and single-layer $WTe_2$ devices that may evade degradation and thus preserve their intrinsic properties and high performance.

## 4. Conclusion

In summary, we have investigated degradation of 1L, 2L and 3L $WTe_2$ using optical characterization (Raman spectroscopy) and materials science surface analytical technicques (XPS and AES). We find relatively easy and fast degradation in single-layer $WTe_2$ (less than 13min for complete oxidation), which is much faster compared with that of many other 2D materials. The main driving force for degradation is oxidation of $WTe_2$ into $WO_x$ and $TeO_2$ on its surface, which is a self-limiting process. Such unique degradation and oxidation behaviors in $WTe_2$ may result from low energy barrier for oxidation. Our results shed light on the mechanisms of $WTe_2$ degradation and pave the way for pursuing high-quality $WTe_2$ single- and few-layer devices, such BN encapsulation, polymer film mixing and refreshing device surfaces using gentle, sequential $Ar^+$ ion plasma etching.

## 5. Experimental Section

Optical Characterization**:** Raman spectroscopy is performed using a customized micro-Raman system. Raman spectra of $WTe_2$ flakes are measured both in vacuum ($p$≈20mTorr) and ambient conditions. A 532nm green laser is focused using a 50× microscope objective, and laser power is limited below ~60μW to avoid laser heating induced crystal modification. Subsequently scattered light from the $WTe_2$ crystal is transmitted to a spectrometer (Horiba iHR550) with a 2400g/mm grating, and recorded by a liquid-nitrogen-cooled CCD. All the measurements are conducted with all samples held at room temperature.

Surface Analysis**:** XPS (PHI Versaprobe 5000 Scanning X-Ray Photoelectron Spectrometer) is performed with binding energies reference to adventitious carbon at 284.6 eV. A spot size of X-ray beam is reduced to 20 μm using an aperture. AES (PHI 680 Scanning Auger Microprobe) is performed with beam voltage of 10 kV. Etching in XPS and AES is performed using $Ar^+$ ions with 3 kV acceleration. The etching rate is first experimentally calibrated by calculating the time of Etching 100nm $Ta_2O_5$ film on Te substrates, which is a conventional calibration standard. After that it is converted to etching rate of mixture of $WO_3$ and $TeO_2$ by comparing etching yields of $WO_3$ (2.75 atom/ion), $TeO_2$ (2.44 atom/ion) and $Ta_2O_5$ (3.12 atom/ion). Etching yields of materials are estimated using SRIM simulation. Since $WO_x$ and $TeO_2$ on the degraded $WTe_2$ are native oxides with weak bonding, actual etching rate may be slightly larger than this estimation.

## Acknowledgements

We thank the support from Case School of Engineering, National Academy of Engineering (NAE) Grainger Foundation Frontier of Engineering (FOE) Award (FOE2013-005), National Science Foundation CAREER Award (Grant ECCS-1454570). Work at Tulane is supported by the DOE under grant DE-SC0014208 (for material synthesis) and the Louisiana Board of Regents under grant LEQSF (2015-18)-RD-A-23.